\documentclass[10pt]{article}

\begin{document}

\begin{center}
\bf{Spectroscopy of a single semiconductor quantum dot
at negative and positive discrete charge states}\\
\vspace{4mm}
\em{D.V.~Regelman $^{1}$ \footnote{e-mail: dany@ssrc.technion.ac.il},
	E.~Dekel $^1$,
         D.~Gershoni $^1$, 
	E.Ehrenfreund $^1$,
          W.V.~Schoenfeld $^2$, 
          P.M.~Petroff $^2$}\\
\vspace{2mm}
{$^1$ Solid State Institute and Physics Department,
 Technion- Israel Institute of Technology, 
  Haifa, 32000, Israel}\\
\vspace{2mm}
{$^2$ Materials Department, University of California, Santa Barbara, CA
93106 USA}
\end{center}

\vspace{3mm}

\begin{abstract}
We study optically single self-assembled quantum dots embedded within the wide quantum well of a mixed type quantum structure. We compare the steady state and pulsed photoluminescence spectra of these dots to those of previously
studied "regular" dots. We unambiguously identify experimentally emission from various discrete charge state of the dots. 
We provide means for optically tune the charge state of the dot, both negatively and {\bf positively}. 
Our observations are used to accurately determine the asymmetry between the quantum dots' confined electron and hole envelope wavefunctions. 

 \end{abstract}


\section{Introduction}
Optical studies of semiconductor quantum dots (QDs)
have been a subject of very intensive recent investigations.
It has been experimentally and theoretically established that the number of carriers which occupy a photoexcited QD greatly affect its 
photoluminescence (PL) spectrum.\cite{gammonNATURE}
In spite of its neutral nature, optical spectroscopy has very recently proved to be a useful means 
for investigating and preparing charged QD systems.\cite{karraiNATURE,kaponPRL}
We report here on continuous wave (cw) and pulsed optical PL spectroscopy
of single self-assembled QDs (SAQDs) embedded within a mixed type quantum
 well (QW) structure.\cite{winstonAPL} This specific design, which facilitates charge separation 
by optical means,\cite{winstonSCIENCE} is used here to tune the charge state of the QD under study.

\section{Experiment}

 Two samples were studied. Sample A consists of a layer of low 
density In(Ga)As SAQDs, embedded within the wider of a two mixed type coupled GaAs QWs, separated by a thin 
AlAs barrier layer.\cite{winstonAPL}. 
Sample B, which is used here as a control sample, consists 
of a similarly prepared SAQDs layer embedded within a thick 
layer of GaAs.\cite{edPRB}

We spatially, spectrally and temporally resolve the PL emission from single SAQDs in both samples using a variable temperature confocal microscope setup.
The setup is described in detail elsewhere.\cite{erezPRB}

In Fig. 1a (1b) we present the PL spectra from sample A(B) for various cw excitation powers 
at photon energy of 1.75 eV. In Fig. 1c (1d) we present the temporally integrated PL spectra from sample A(B) for various 
picosecond pulsed excitation powers by the same photon energy.
The repetition rate of our Ti:sapphire laser which was used for pulsed excitation is 78MHz.

In Fig. 1d (1e) we present for comparison model simulations of the pulsed excitation PL spectra for sample A(B) 
as explained below.

\section{Discussion}

In previous works[6-8]
we have established that 
the evolution of the cw PL emission from an optically excited SAQD is due to the recombination of increasingly 
higher orders of neutral multiexcitons. 
Characteristically, with the increase in power, satellite spectral lines
appear to the lower energy side of the first observed line ($X_0$ in Fig. 1b) 
and higher energy spectral group of lines ($B$) emerges 
above the first observed group ($A$).
This spectral red shift is explained in terms of the exchange energies between the increasing numbers of 
electron-hole pairs within the neutral SAQD.
At yet higher excitation powers, all the observed discrete PL lines at their appearance order, 
undergo a cycle in which their PL intensity first increases, then reaches maximum and saturates, 
and eventually significantly weakens. Consequently, the various groups of spectral lines (A and B in Fig. 1b) seem to
be "red shifted" with the increase in excitation power.
This behavior is best described by a set of coupled rate equations\cite{edPRB}, 
which give the probabilities of finding the photoexcited QD occupied by a given number of 
e-h pairs (multiexciton order $N_x$). At high excitation power the 
probability to find the QD with small $N_x$ 
vanishes and thereby the PL lines observed at low powers disappear.
This is no longer true when pulse excitation is used. In this situation
with the increase in excitation power, after reaching saturation,
the intensity of the PL of the various spectral lines remains constant
and does not decrease with further increase in the excitation power (Fig. 1c). 
In this case the radiative recombination process is sequential.
Therefore, all the multiexcitons which are smaller than the initially photogenerated 
number of e-h pairs ($<N_x>$) contribute to the temporally
integrated PL spectrum. This is true, as long as the pulse repetition rate 
is slow such that all the photogenerated pairs recombine before the next pulse arrives.

In sample A, the SAQDs are initially charged with electrons. These electrons accumulate in the SAQDs due to the special design of the sample which
facilitates efficient hoping transport from residual N-type impurities. 
With the increase in the excitation power the SAQDs are first photodepleted, 
since the positively charged traps efficiently capture photogenerated electrons
while holes diffuse preferentially to the SAQDs.\cite{kaponPRL}
The recombination of an e-h pair in the presence of a decreasing number
of electrons within the SAQD reveals itself in a series of small lines
to the lower energy side of the PL line due to recombination in a neutral
SAQD (line $X_0$ in Fig 1a ). The higher the excitation power is, the less charged
the dot is and consequently the PL line is closer to the $X_0$ line.
The intensity ratio between the saturated PL lines from the dot in the 
presence of charge and that from the neutral SAQD is roughly 
an order of magnitude. This ratio must be equal to the ratio between 
the rate by which the photexcited electrons leave the impurities and 
the SAQD intrinsic radiative
rate. In a recent work we showed that the later amounts to 5 ns.\cite{edPRB}. 
It follows that the lifetime associated with the trapped photoexcited electrons
is few times longer than the time between sequential pulses. Therefore, under
pulsed excitation, PL lines from charged states saturate with increasing
the excitation power and lose their intensity as the power is further increased.
At the same time the PL from neutral states, saturates and remains constant
(see Fig. 1d). This behavior provides an unambiguous experimental tool for the 
distinction between the two PL processes. 

Comparison between the PL spectra of sample A and sample B reveals yet another
distinctive difference. 
With increasing the excitation power satellite PL lines appear to the i
higher energy side of the neutral $X_0$ line of sample A. 
These lines can now be unambiguously identified as resulting from 
recombination from QDs that are {\bf positively} charged.

The fact that PL from negatively (positively) charged QD is lower (higher) 
in energy than the PL from neutral QD is reported here, to the best of our knowledge, for the first time.
This phenomenon can be straightforwardly explained by solving the 
confined few carriers problem,\cite{BD} with different
confining potentials, for electrons and  for holes.
The experimental results that we obtained are best fitted by our model,[6-8]
 as demonstrated in Fig. 1(e,f), when a 25\% smaller hole envelope
wavefunction is assumed.
Intuitively it means that the energy associated with the increase Coulombic 
repulsion due to the additional positive holes is greater than the 
additional exchange energy and the energy 
associated with the electrons-hole attraction.
\section{Summary}

We report on the first observation of photoluminescence from
a positively charged single quantum dot. Our experimental observations
strongly suggest that in self assembled quantum dots holes are better confined
than electrons.

{\bf Acknowledgments}: The research was supported by the US-Israel
Binational Science Foundation (453/97),
by the Israel Science Foundation founded by the Israel Academy of
Sciences, and by the Fund for the Promotion of Research at the Technion.

\begin{figure}
\caption{a(b) [c(d)] PL spectra from sample A(B) for various cw [pulsed] excitation powers.
e(f) model simulations of the pulsed excitation PL spectra from sample
A(B) for various initial avarage numbers
of e-h pairs $<N_x>$. }
\label{fig:1}
\end{figure}

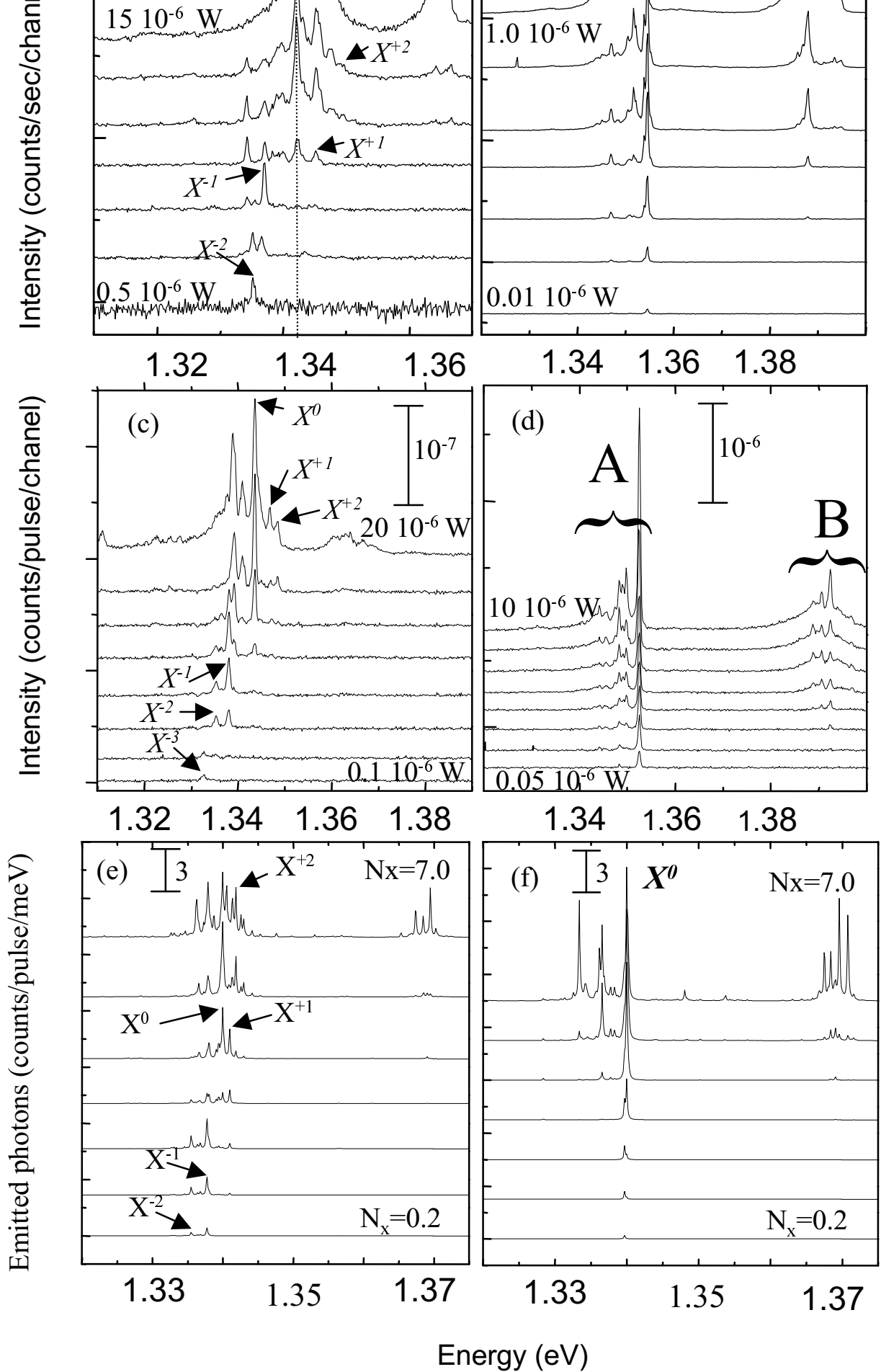
\end{document}